# Modeling and Analysis of Abnormality Detection in Biomolecular Nano-Networks

**Siavash Ghavami[1, 2], Farshad Lahouti[*1], Ali Masoudi-Nejad[2]**

1. Center for Wireless Multimedia Communications (WMC), School of Electrical & Computer Engineering, College of Engineering, University of Tehran, Tehran, Iran

2. Laboratory of Systems Biology and Bioinformatics (LBB), Institute of Biochemistry and Biophysics, University of Tehran, Iran.

Siavash Ghavami       : s.ghavami@ut.ac.ir
Farshad Lahouti       : lahouti@ut.ac.ir
Ali Masoudi-Nejad     : amasoudin@ibb.ut.ac.ir

**\*Corresponding Author**
Farshad Lahouti
Center for Wireless Multimedia Communications
School of Electrical & Computer Engineering, College of Engineering, University of Tehran, Tehran, Iran
Tel: +98(21) 82084314
E-mail: lahouti@ut.ac.ir

*Abstract*

A scheme for detection of abnormality in molecular nano-networks is proposed. This is motivated by the fact that early diagnosis, classification and detection of diseases such as cancer play a crucial role in their successful treatment. The proposed nano-abnormality detection scheme (NADS) comprises of a two-tier network of sensor nano-machines (SNMs) in the first tier and a data gathering node (DGN) at the sink. The SNMs detect the presence of competitor cells as abnormality that is captured by variations in parameters of a nano-communications channel. In the second step, the SNMs transmit micro-scale messages over a noisy micro communications channel (MCC) to the DGN, where a decision is made upon fusing the received signals. The detection performance of each SNM is analyzed by setting up a Neyman-Pearson test. Next, taking into account the effect of the MCC, the overall performance of the proposed NADS is quantified in terms of probabilities of misdetection and false alarm. A design problem is formulated, when the optimized concentration of SNMs in a sample is obtained for a high probability of detection and a limited probability of false alarm.

# 1. Introduction

Cancer is the leading cause of death in economically developed countries and the second leading cause of death in developing countries [1]. In fact, more than 7.6 million people lose their lives each year in the world due to the various forms of cancer [2]. Despite the advances made by significant innovations in the technology in the past decade, successful treatment of cancer is still a distant goal in general. Accurate diagnosis, classification and early detection of cancer are vital in overcoming the disease [3].

The abnormality detection (AD) process based on molecular diagnosis can be used as an effective method for early cancer detection. Systems in the molecular environment are to distinguish two molecules *A* and *B* [4][5]. The decisions are made using a recognizer molecule $a$. For example, molecule $a$ is an antigen and molecule *A* is a pathogen that is to be diagnosed by molecule $a$, while the molecule *B* which is similar to *A*, is normally found in the body. The design of optimized molecular recognizers is studied in the biochemical noisy environment using a Bayesian cost in [4].

A diagnosis process based on protein identification and considering reaction between genes and antigens with magnetic nano-sensors is investigated in [6][7]. This type of nano-sensors can detect a nano-scale message by generating a specific signal in micro-scale. In [8], the detection of abnormalities is examined using biomarkers in the blood as a detection feature. To identify breast cancer based on a breath test, a proof-of-concept system has been developed using a tailor-made nano-scale artificial nose [9].

In this paper, a nano-abnormality detection scheme (NADS) is proposed for the detection of nano-scale abnormality in a biomolecular environment using a two tier decision making process. The abnormality is due to the existence of competitor cells in the said environment. The NADS includes a set of SNMs for the detection of a nano-scale abnormality over a nano-communication channel (NCC) using an appropriately designed detection feature. Then, the SNMs communicate their observations over a noisy micro communication channel (MCC) to a data gathering node (DGN) using micro scale messages (MSMs). Fusing the collected signals, the DGN makes a decision and may alarm an abnormality as necessary. The analysis of SNM performance over the NCC is set up as a Neyman-Pearson test, which quantifies the probabilities of false alarm and misdetection. Finally, incorporating the effect of MCC, the total detection performance of NADS at the DGN is analyzed. The presented analysis is then used to obtain the optimized concentration of SNMs in the sample tissue for a prescribed high probability of detecting the abnormality and a bounded false alarm probability.

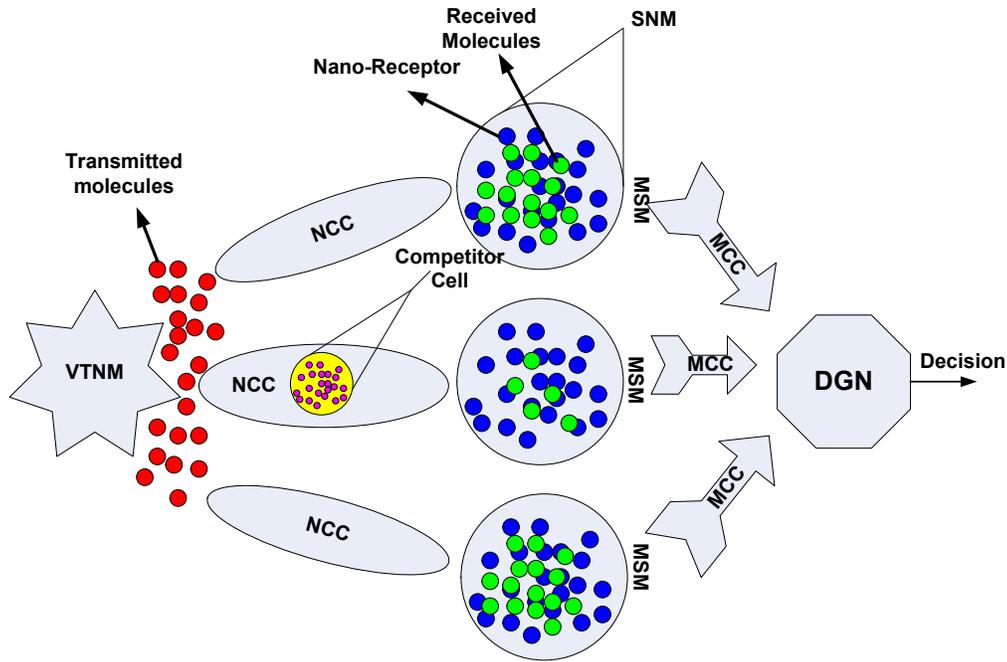

Fig. 1. System model for NADS.

Certain schemes for cancer detection at the nano-scale, may be cast into the proposed two-tier network model of the NADS. Nano-sized magnetic resonance imaging (MRI) contrast agents is used for intraoperative imaging in the context of neuro-oncological interventions [10][11]. In this scheme, detection in NCC is done using, Gadolinium-based nano-particle [12], Ironoxide-based nano-particles or multiple-mode imaging contrast nano-agents [13]-[18]. Moreover, detection over the MCC is accomplished using a combination of MRI and biological targeting [19] or optical detection [11][19][20]. As another scheme, nano-scale field-effect bio-transistors can be considered, in which silicon nano-wires [21][22] detect an abnormal target in the NCC due to the molecular binding on their surface. This in turn affects the conductance of nano-wires, which is reported over the MCC.

The outline of this paper is as follows. In Section II, the preliminaries and problem statement are presented. The hypothesis test at an SNM for AD in nano-scale is derived in Section III. In Section IV, communications over the MCC is studied. Section V presents the analysis of the NADS performance and the associated numerical results. Concluding remarks are made in Section VI.

## 2. Preliminaries and Problem Statement

In this section, the setup of AD system and the problem statement of NADS are described. The detection process includes two parts. In the first part, each SNM detects the detection feature in nano-scale and generates the MSM; in the second part, the DGN collects the transmitted MSMs.

Fig. 1 shows the schematic structure of NADS. The molecular environment is modeled by the NCC. In the healthy setting, no abnormality (here competitor cell) exists in the molecular environment and vice versa. The molecular competitor changes the rate of binding between the molecules and the nano-receptors on the SNM. This is reflected in the NCC model [23][24]. The MSM is then generated by the SNM as it detects this variation. The DGN collects the MSMs over the noisy MCCs and decides and declares the presence or the absence of the abnormality. The MCC is considered an additive white Gaussian noise (AWGN) channel. Below, we continue with a detailed description of the NCC model and the detection feature.

### 2.1. Nano-Communication Channel

The NCC characterizes chemical reactions in the molecular environment. We consider a set of SNMs, which act as molecular receivers, injected into the biological tissue for test. We assume that the existing molecules in the molecular environment react with the receptors on SNMs. The molecules are assumed to be propagated by a virtual transmitter nano-machine (VTNM) following a periodic square pulse propagation pattern, i.e., during time interval $t_H$ a molecular pulse is emitted with concentration $L(t) = L_{ex}$, and otherwise we have $L(t) = 0$ [25]. At the SNM receptor, if at least $S$ molecules are delivered within time interval $t_H$, a molecular bit $A$ is received. If the number of molecules delivered within $t_H$ is less than $S$, the SNM receives a molecular bit $B$. If the VTNM transmits a molecular pulse, the number of received molecules during the time $t_H$ is quantified by

$$N_A = \int_0^{t_H} C(t) dt. \tag{1}$$

in which $C(t)$ denotes the concentration of bound receptors at the SNM in $\mu mol/liter$ and is given by

$$C(t) = C(\infty)\left(1 - \exp\left(-t(\kappa_{-1} + \kappa_1 L_{ex})\right)\right), \tag{2}$$

and $C(\infty) = \kappa_1 L_{ex} N / (\kappa_{-1} + \kappa_1 L_{ex})$ is the steady state concentration of bound receptors [24]. The related parameters are described in Table 1. The parameters $\kappa_1$ and $\kappa_{-1}$ respectively are the binding and the release rates related to reactions of $A + R \xrightarrow{\kappa_1} D$ and $A + R \xleftarrow{\kappa_{-1}} D$, where $R$ and $D$ respectively, denote nano-receptors on the SNMs and the product of reaction between $A$ and $R$. It is evident in (2) that $C(t)$ increases exponentially over time within the pulse period. When the pulse duration ends, $C(t)$ reduces as

$$C(t) = C(t_0) \exp\left(-\kappa_{-1}(t - t_0)\right) \quad for \quad t > t_0. \tag{3}$$

As a result, at the SNM and over the subsequent time interval, this previous pulse is reflected as follows

$$N'_A = \int_0^{t_H} N_A \exp(-\kappa_{-1} t) dt. \tag{4}$$

Furthermore, if the VTNM does not propagate a molecular pulse in the current transmission interval, the corresponding number of received molecules at the SNM is assumed zero, $N_B = 0$, and similarly the remaining number of molecules in the subsequent time interval is zero, $N'_B = 0$. The NCC is then considered as a binary channel with finite state, where the channel state is determined based on the received molecular bit ($A$ or $B$) in the previous time interval. If the molecular bit $A(B)$ is received in the previous time interval (denoted by $A_R^P$ ($B_R^P$)), the transition probability of the NCC is denoted by $p_{ij}^{A_R^P}$ ($p_{ij}^{B_R^P}$), $i, j \in \{A, B\}$. These probabilities are bounded using the Markov (aka Chebyshev) inequality as follows [24][26]

$$p_{AA}^{A_R^P} = p\left(N_A^C > S \big| A_R^P\right) < E\left[N_A^C \big| A_R^P\right] \big/ S, \tag{5}$$

$$p_{AB}^{A_R^P} = 1 - p\left(N_A^C > S \big| A_R^P\right) > 1 - E\left[N_A^C \big| A_R^P\right] \big/ S, \tag{6}$$

$$p_{BA}^{A_R^P} = 1 - p\left(N_B^C < S \big| A_R^P\right) < E\left[N_B^C \big| A_R^P\right] \big/ S, \tag{7}$$

$$p_{BB}^{A_R^P} = p\left(N_B^C \leq S \big| A_R^P\right) > 1 - E\left[N_B^C \big| A_R^P\right] \big/ S. \tag{8}$$

In a similar way, $p_{ij}^{B_R^P}$ is derived. In (5)-(8), $N_A^C$ ($N_B^C$) denotes the average number of received molecules when molecular bit $A$ ($B$) is transmitted. We have

$$E\left[N_A^C \big| A_R^P\right] = N_A + N_A', \tag{9}$$

$$E\left[N_A^C \big| B_R^P\right] = N_A + N_{B'}', \tag{10}$$

$$E\left[N_B^C \big| A_R^P\right] = N_B + N_A', \tag{11}$$

$$E\left[N_B^C \big| B_R^P\right] = N_B', \tag{12}$$

The transition probability matrix of the NCC using the described parameters and considering $0 \leq p_{ij} \leq 1$, $i, j \in \{A, B\}$ and the channel state is given by

$$\mathbf{P}^{A_R^P} = \begin{bmatrix} p_{AA}^{A_R^P} & p_{AB}^{A_R^P} \\ p_{BA}^{A_R^P} & p_{BB}^{A_R^P} \end{bmatrix} = \max\left( \min\left( \begin{bmatrix} E\left[N_A^C \middle| A_R^P\right]/S & 1 - E\left[N_A^C \middle| A_R^P\right]/S \\ E\left[N_B^C \middle| A_R^P\right]/S & 1 - E\left[N_B^C \middle| A_R^P\right]/S \end{bmatrix}, \begin{bmatrix} 1 & 1 \\ 1 & 1 \end{bmatrix} \right), \begin{bmatrix} 0 & 0 \\ 0 & 0 \end{bmatrix} \right), \qquad (13)$$

$$\mathbf{P}^{B_R^P} = \begin{bmatrix} p_{AA}^{B_R^P} & p_{AB}^{B_R^P} \\ p_{BA}^{B_R^P} & p_{BB}^{B_R^P} \end{bmatrix} = \max\left( \min\left( \begin{bmatrix} E\left[N_A^C \middle| B_R^P\right]/S & 1 - E\left[N_A^C \middle| B_R^P\right]/S \\ E\left[N_B^C \middle| B_R^P\right]/S & 1 - E\left[N_B^C \middle| B_R^P\right]/S \end{bmatrix}, \begin{bmatrix} 1 & 1 \\ 1 & 1 \end{bmatrix} \right), \begin{bmatrix} 0 & 0 \\ 0 & 0 \end{bmatrix} \right). \qquad (14)$$

Table 1. Effective parameters in transition probability of the NCC [24]

| Parameter | Parameter Description | Unit |
|---|---|---|
| $T$ | Temperature | *Kelvin* |
| $\alpha$ | Distance between nano-transmitter and SNM | *meter* |
| $N$ | Number of receptors on each SNM | $\mu mol$ |
| $t_H$ | Pulse duration | *sec* |
| $L_{ex}$ | Molecules concentration | $\mu mol/(liter.\sec)$ |
| $\kappa_1$ | Binding rate | $\mu mol/(liter.\sec)$ |
| $\kappa_{-1}$ | Release rate | $\mu mol/(liter.\sec)$ |
| $k_B$ | Boltzman constant | *Joule/Kelvin* |
| $S$ | If the SNM can receive $S$ molecules, it infers that the VTNM emitted the molecule $A$ during $t_H$. | $\mu mol/(liter.\sec)$ |
| $N_A^C$ ($N_B^C$) | The number of received molecules, when VTNM sends the molecule $A(B)$ during time $t_H$. | $\mu mol$ |
| $P_A$ | Probability of transmission of molecule $A$ by the VTNM. | |
| $P_A'$ | Probability of receiving molecular bit $A$ at the SNM. | |

Table 1 summarizes the parameters affecting the NCC transition probabilities.

*2.2. Detection Feature*

The biochemical activities of competitor cells, e.g., cancer cells, affect the molecular environment and change its parameters [4]. We model this as an abnormality in the molecular environment, which is to be detected as early as possible. The competitor cells can be considered as competitor nano-machines (CNMs) for the SNMs. The presence of CNM affects the transition probability of the NCC in (13) and (14). For example, the CNM can react with transmitted molecules by the VTNM. This reduces the concentration of transmitted molecules $L_{ex}$, and hence, changes the transition probability of the NCC. This variation in transition probability can be used for modeling of protein identification for early cancer detection in nano-scale [27]. Alternatively, the CNM may devitalize the receptors on the SNMs, change $\kappa_{-1}$ and $\kappa_1$ on the SNM by a biochemical reaction or vary the temperature of nano-receptors on SNMs.

For the NCC, with given parameters in Table 1 and a given size of sample tissue, the number of excitation pulses per unit time from VTNMs influences the probability of receiving a given molecular bit in the SNM nano-receptor. Hence, we consider the average number of molecular bits $A$ received at the SNMs per unit time, $NP_R$, as the detection feature. The value of this parameter in a healthy setting, $NP_H$, is expected to be measured using biological experiments. Still, to reveal the dependency of this detection feature on the characteristics of the biomolecular environment, we may consider the presented model for the NCC and quantify the average number of molecular bits received per unit time in the healthy setting as follows (see Appendix A for proof)

$$\begin{aligned}
NP_H &= \min\left(\frac{L_{ex,V}}{S-N'_A}, N\right) \frac{p_{AA}^{A^P_R} P_A \left(p_{AA}^{B^P_R} P_A + p_{BA}^{B^P_R}(1-P_A)\right)}{1 - p_{AA}^{A^P_R} P_A - p_{BA}^{A^P_R}(1-P_A) + p_{AA}^{B^P_R} P_A + p_{BA}^{B^P_R}(1-P_A)} \\
&+ \min\left(\frac{L_{ex,V}}{S-N'_B}, N\right) \frac{p_{AA}^{B^P_R} P_A \left(p_{AB}^{B^P_R} P_A + p_{BB}^{B^P_R}(1-P_A)\right)}{1 - p_{AB}^{A^P_R} P_A - p_{BB}^{A^P_R}(1-P_A) + p_{AB}^{B^P_R} P_A + p_{BB}^{B^P_R}(1-P_A)} \\
&+ \min\left(\frac{L_{ex,V}}{S-N'_A}, N\right) \frac{p_{BA}^{A^P_R}(1-P_A)\left(p_{AA}^{B^P_R} P_A + p_{BA}^{B^P_R}(1-P_A)\right)}{1 - p_{AA}^{A^P_R} P_A - p_{BA}^{A^P_R}(1-P_A) + p_{AA}^{B^P_R} P_A + p_{BA}^{B^P_R}(1-P_A)} \\
&+ \min\left(\frac{L_{ex,V}}{S-N'_B}, N\right) \frac{p_{BA}^{B^P_R}(1-P_A)\left(p_{AB}^{B^P_R} P_A + p_{BB}^{B^P_R}(1-P_A)\right)}{1 - p_{AB}^{A^P_R} P_A - p_{BB}^{A^P_R}(1-P_A) + p_{AB}^{B^P_R} P_A + p_{BB}^{B^P_R}(1-P_A)}
\end{aligned} \quad (15)$$

When $NP_R$ deviates from $NP_H$, an abnormality or the existence of competitor cells is detected.

*2.3. Problem Statement*

Here, we consider a design problem to determine the minimum required concentration of SNMs, $M$, in the sample size of $V$ for a reliable NADS. The SNMs in general could be expensive and/or are chemical compounds which could affect the molecular environment or cause side effects if used in vivo. This motivates us to use them in smallest amount as possible. A reliable NADS is interpreted as one which provides a sufficiently high probability of detection (alarm), when the abnormality exists and a sufficiently small probability of (false) alarm, when the abnormality in fact does not exist in the molecular environment. Hence, the desired optimization problem is formulated as follows

$$\min M$$
$$\text{s.t.}$$
$$P_D > \xi, P_F < \gamma. \tag{16}$$

where $P_D$ and $P_F$ are the detection and the false alarm probabilities of NADS on detecting the abnormality in the molecular environment, respectively. Also, $\xi < 1$ is a constant close to unity and $\gamma$ is a positive constant close to zero.

## 3. Hypothesis Test for AD in the NCC

In this section, a hypothesis test is formulated for the detection of competitor cells in the molecular environment. This test determines the functionality of SNMs in the nano-scale. We derive a threshold level for generating the MSM by the SNM, which alarms the existence of competitor cells. To this end, for a bounded false alarm probability, a closed-form expression is derived for the probability of each SNM to detect the presence of different types of competitor cells. The associated hypothesis test is defined as

$$\begin{cases} H_0, & NP_R = NP_H \\ H_1, & NP_R \neq NP_H. \end{cases} \tag{17}$$

In line with [28], the SNMs observations, $y_i$s, are assumed Poisson distributed with mean $NP_R$. Hence, the probability mass function (PMF) of $y_i$s is given by

$$P(y_i = q) = \exp(-NP_R) NP_R^q / q!, \tag{18}$$

where ! denotes the factorial function. If an SNM makes $n$ independent observations for the detection, the probability of $\mathbf{y}$ given $NP_R$ is given by

$$P(\mathbf{y}|NP_R) = P(y_1, \ldots, y_n | NP_R) = \prod_{i=1}^{n} P(y_i | NP_R)$$
$$= \frac{\exp(-nNP_R) NP_R^{\sum_{i=1}^{n} y_i}}{\prod_{i=1}^{n} y_i!}, \tag{19}$$

and hence

$$\log P(\mathbf{y}|NP_R) = -n NP_R + \sum_{i=1}^{n} y_i \log_e NP_R - \sum_{i=1}^{n} \log_e y_i!. \tag{20}$$

To maximize (20), we set its derivative with respect to $NP_R$ to zero. We have

$$\widehat{NP_R} = \arg\max_{NP_R} \left( P(\mathbf{y}|NP_R) \right) = \frac{1}{n} \sum_{i=1}^{n} y_i \tag{21}$$

For deriving the decision rule of the Neyman-Pearson as in the hypothesis test of (17), we employ the generalized likelihood ratio test (GLRT) in the next theorem.

*Theorem* 1. Consider an SNM with $n$ Poisson observations over the NCC. For the hypothesis test in (17), the decision threshold with limited false alarm probability, $P_F^{NCC} < \eta_1$, are given by

$$\begin{cases} H_0, & \dfrac{1}{n}\sum_{i=1}^{n} y_i \left( \log_e \left( \dfrac{1}{n} \sum_{i=1}^{n} \dfrac{y_i}{NP_H} \right) - 1 \right) < \dfrac{\log(\tau)}{n} - NP_H \\[2mm] H_1, & \dfrac{1}{n}\sum_{i=1}^{n} y_i \left( \log_e \left( \dfrac{1}{n} \sum_{i=1}^{n} \dfrac{y_i}{NP_H} \right) - 1 \right) > \dfrac{\log(\tau)}{n} - NP_H. \end{cases} \tag{22}$$

*Proof.* See Appendix B.□

One sees in (22) that the decision threshold is a nonlinear function of the measurements $y_i$. In fact, its implementation in an SNM with limited complexity may be challenging. In addition, the threshold in (22) does not enable tractable analytical assessment of $P_D^{NCC}$ and $P_F^{NCC}$. Hence, motivated by our experiments on (22), a new simplified decision threshold is proposed based on the law of large numbers (LLN). According to the LLN, the average of a large number of observations of a random variable tends to its mean or expectation.

The progress of cancer is modeled with parameter $k$, which corresponds to a given value of parameter $k \neq 1$. For a healthy molecular environment, we have $k = 1$. Hence, for a given value of $k$, we may consider

$$NP_R = k\, NP_H. \tag{23}$$

The next theorem presents the resulting decision threshold for the hypothesis test and the detection probability.

*Theorem 2.* Consider an SNM with $n$ Poisson observations over the NCC. For the hypothesis test in (17), the simplified decision threshold and detection probability, $P_D^{NCC}$, with limited false alarm probability, $P_F^{NCC} < \eta_1$, are given by

$$\begin{cases} H_0, & \max\{(NP_H - \tau''), 0\} \leq \dfrac{1}{n}\sum_{i=1}^{n} y_i \leq NP_H + \tau'' \\ H_1, & \begin{cases} \dfrac{1}{n}\sum_{i=1}^{n} y_i > NP_H + \tau'' \\ \dfrac{1}{n}\sum_{i=1}^{n} y_i < \max\{(NP_H - \tau''), 0\}, \end{cases} \end{cases} \tag{24}$$

$$P_D^{NCC} = 1 + \frac{\Gamma\left(\lfloor \max\{n(NP_H - \tau''), 0\} - 1 \rfloor + 1, nkNP_H\right)}{\lfloor \max\{n(NP_H - \tau''), 0\} - 1 \rfloor !} - \frac{\Gamma\left(\lfloor n(NP_H + \tau'') \rfloor + 1, nkNP_H\right)}{\lfloor n(NP_H + \tau'') \rfloor !} \tag{25}$$

where $\Gamma(.,.)$ and $\lfloor . \rfloor$ are respectively the incomplete Gamma function and the floor operator. Also, $\tau'' \in \mathbb{R}$ is the smallest number which satisfies the following

$$\frac{\Gamma\left(\lfloor n(NP_H + \tau'') \rfloor + 1, nNP_H\right)}{\lfloor n(NP_H + \tau'') \rfloor!} - \frac{\Gamma\left(\lfloor \max\{n(NP_H - \tau''), 0\} - 1 \rfloor + 1, nNP_H\right)}{\lfloor \max\{n(NP_H - \tau''), 0\} - 1 \rfloor!} \geq 1 - \eta_1 \qquad (26)$$

*Proof.* See Appendix C. □

Using (25), the probability of misdetection is given by

$$P_M^{NCC} = 1 - P_D^{NCC}. \qquad (27)$$

## 4. Communications over the MCC

The DGN receives the MSMs from the SNMs over the MCC and declares the existence or the absence of the competitor cells in the NCC. It is assumed that this message has two alphabets. If the $j^{th}$ SNM detects the competitor cells, it generates the message $X_j = G$. In this section, we derive the probability of detecting an MSM at the DGN. If the competitor cells are present in the molecular environment, noting the possible erroneous detection of the SNM, it generates the following MSM

$$X_j = \begin{cases} 0, & P_M^{NCC} \\ G, & 1 - P_M^{NCC}. \end{cases} \qquad (28)$$

If the competitor cells are not present in the NCC, we have

$$X_j = \begin{cases} 0, & 1 - P_F^{NCC} \\ G, & P_F^{NCC}, \end{cases} \qquad (29)$$

The signal received at the DGN through the AWGN MCC is then given by

$$U = \sum_{j=1}^{M} X_j + Z, \qquad (30)$$

where, $Z \sim \mathcal{N}(0, \sigma_{MCC}^2)$, and $\sigma_{MCC}^2$ is the MCC noise variance. We set up the following hypothesis test at the DGN,

$$\begin{cases} H_0, & \sum_{j=1}^{M} X_j < G \\ H_1, & \sum_{j=1}^{M} X_j \geq G. \end{cases} \quad (31)$$

This fusion rule is known as the OR-rule [29], which means that if one or more SNMs alarm the presence of abnormality in the biomolecular environment, the DGN alarms the same. The hypothesis $H_1$ is declared if at least one of the SNMs transmits the MSM $G$.

The next proposition presents the decision threshold and the resulting performance of the decision in (31) at the DGN.

*Proposition 1.* The decision threshold and the resulting probabilities of misdetection and false alarm for the hypothesis test in (31) at the DGN over the AWGN MCC are given by

$$\begin{cases} H_0, & U < G/2 \\ H_1, & U \geq G/2. \end{cases} \quad (32)$$

$$P_M^{MCC} = P_F^{MCC} = Q(G/2\sigma_{MCC}). \quad (33)$$

*Proof.* See Appendix D.

## 5. NADS Performance Assessment

In this Section, first performance analysis of NADS is done. Then this analytical results are evaluated numerically in Subsection B.

*A. Analytical Performance Evaluation*

In this Subsection, analytical expressions for probabilities of detection and false alarm of the NADS are derived. To this end, an equivalent channel is obtained between the VTNMs. and the DGN by cascading the NCC and the MCC. The transition probabilities of the two channels are respectively derived in Theorems 2 and Proposition 1. The next proposition presents the performance of the NADS.

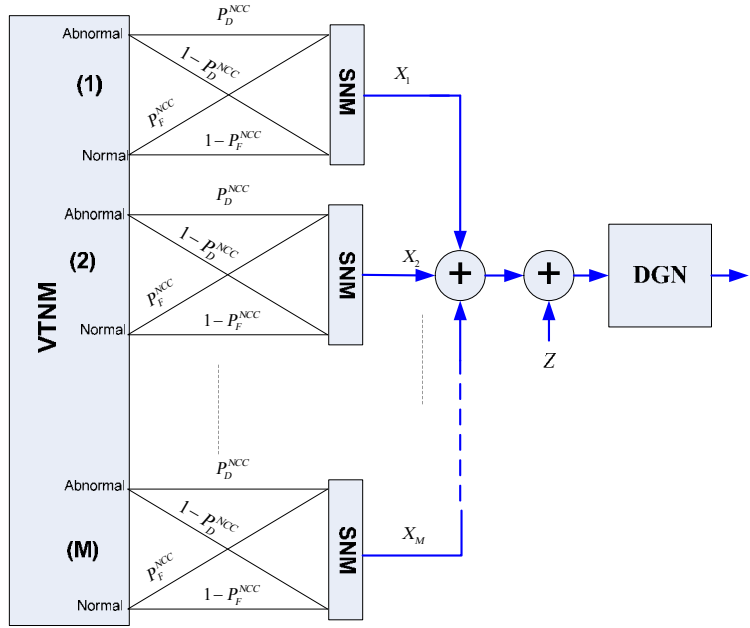

Fig. 2. Equivalent channel between VTNM and SNMs, and between SNMs and DGN.

*Proposition 2.* For the proposed NADS, the detection and false alarm probabilities at the DGN are given by

$$P_D = \left(1 - \left(P_M^{NCC}\right)^M\right)\left(1 - P_M^{MCC}\right) + \left(P_M^{NCC}\right)^M P_F^{MCC}, \qquad (34)$$

$$P_F = \left(1 - \left(1 - P_F^{NCC}\right)^M\right)\left(1 - P_M^{MCC}\right) + \left(1 - P_F^{NCC}\right)^M P_F^{MCC}, \qquad (35)$$

where $P_F^{NCC} = \eta_1$ and $P_M^{NCC}$ and $P_F^{MCC}$ are defined in (27) and (33).

*Proof.* The NADS is composed of a broadcast channel with a common message from the VTNM in the first tier followed by Gaussian multiple access channel and an OR fusion rule [29] in the second tier. Fig. 2 shows a schematic diagram of the NADS. Theorems 2 and Proposition 1 quantify the performance of the two tiers. The NADS performance may be computed by considering two cascaded equivalent channels as shown in Fig. 3, which directly leads to (34) and (35).□

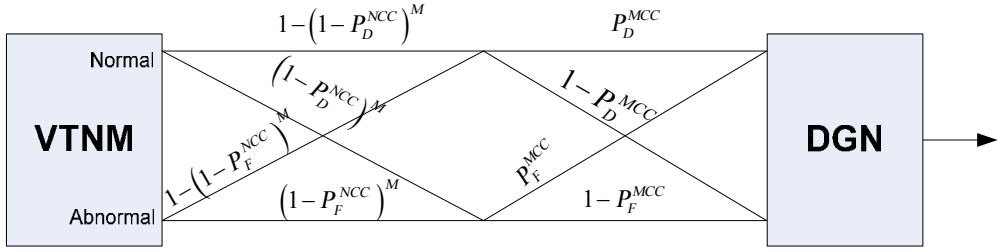

Fig. 3. Equivalent channel model between VTNM and DGN.

In this case, the probability of misdetection for the proposed NADS is given by $P_M = 1 - P_D$. Our extensive experiments reveal that for the overall performance of the NADS, the designed MCC detector in Proposition 1, leads to a lower bound for $P_D$ and a tight upper-bound for $P_F$ (in comparison to those of an optimal detector). In addition, for an MCC with reasonably small error probability, e.g., a high performance medical imaging system, the said bounds approach the optimum solution.

*B. Numerical Results*

In this subsection, numerical results are provided for evaluation of the NADS performance. The parameters of the experiment set up in different figures are presented in Table 2. We start with studying the effect of the characteristics of the SNM, in terms of its observation time and false alarm performance, on the NADS performance. Next, we examine how different levels of abnormality in the molecular environment (NCC) reflect in the NADS performance. Finally, the effect of noise intensity in the second tier of NADS on the overall performance is investigated.

Fig. 4 shows the probability of misdetection, $P_M$, in terms of the concentration of SNMs in the sample size, $M$, for different values of observation time, $n$. One sees that by increasing $M$ and $n$, $P_M$ is reduced. The error floor in this figure is due to the noise of the MCC. Fig. 5 shows the probability of false alarm, $P_F$, in terms of $M$, for the same parameters. It is evident that with $M \leq 10$, $P_F \leq 10^{-5}$ is achieved. This performance result is valid for all values of *n*, since in (35) for the given parameters in Table 2, $P_{MCC}$ is fixed and $P_{NCC}$ holds with equality in (55) and is hence independent of *n*.

The presented results may be used to solve the optimization problem of (16). Using Figs. 4 and 5, for $\xi = 1-10^{-6}$, $\gamma = 10^{-5}$ and the observation time, $n = 9$, the optimized concentration of SNMs, *M*, is obtained as 8 per 1000 $nm^3$.

Figs. 6 and 7 respectively show $P_M$, and $P_F$ of the NADS in terms of *M* for different values of $P_F^{NCC}$. The typical trade-off of false alarm and detection performance of SNM over the NCC is depicted in Fig. 8; as $P_F^{NCC}$ increases $P_M^{NCC}$ reduces and vice versa. Interestingly, $P_F^{NCC}$ affects the overall detection performance of NADS in the same way (Fig. 6), as it directly influences the NADS false alarm performance $P_F$ (Fig. 7).

Fig. 9 shows the probability of misdetection over the NCC as a function of *k*. It is evident that by increasing *k*, $P_M^{NCC}$ reduces. Such a behavior is then reflected in the overall system performance as depicted in Fig. 10.

Figs. 10 and 11 show respectively, $P_M$ and $P_F$ of the NADS in terms of *M* for different values of *k*. One sees that as *k* increases, $P_M$ reduces much faster with *M*. The results indicate that if the competitor cell affects the molecular environment more strongly, its presence is detected more easily. A larger value of *k* in (23), may be interpreted as a disease which has progressed more and hence altered the status of the molecular environment more significantly from a healthy setting. Nevertheless, the detection performance of the NADS is still bounded by the noisy setting of MCC as evident in the error floor of the curves in Fig. 10. As expected and evident in Fig. 11, $P_F$ increases with *M*, and remains the same as *k* changes. This is due to the fact that in (35), for the given parameters in Table 2, $P_{MCC}$ is fixed and $P_{NCC}$ holds with equality in (55) and is hence independent of $k$.

Figs. 12 and 13 show $P_M$ and $P_F$ in terms of $M$ for different values of $\sigma_{MCC}$. For non-small values of $\sigma_{MCC}$, the MCC noise imposes an error floor to the NADS detection performance. It is also evident that $P_F$ increases as $M$ or $\sigma_{MCC}$ increase. However, the performance remains the same, and only a function of $M$, for different but small values of $\sigma_{MCC}$. This is due to the fact that for small $\sigma_{MCC}$, where $P_F^{MCC} \approx 0$ and $P_M^{MCC} \approx 0$, the $P_F$ performance of NADS in (35) dominantly depends on NCC characteristics ($P_F^{NCC}$).

Table 2. Parameters of numerical results in different figures; X:Y:Z denotes the range of parameter as [X,Z] with step size Y. $G=1$, $V=1000\ [nm^3]$, $NP_H = 1$.

| Parameter | $\sigma_{MCC}$ | $\eta_1$ | $n$ | $k$ |
|---|---|---|---|---|
| Figs. 4 and 5 | 0.1 | $10^{-6}$ | 1:2:9 | 2 |
| Figs. 6-8 | 0.1 | $[10^{-4}\ 10^{-3}\ 10^{-2}\ 10^{-1}]$ | 1 | 2 |
| Figs. 9-11 | 0.1 | $10^{-6}$ | 1 | 1.2:0.8:6 |
| Figs. 12 and 13 | 0.03:0.03:0.15 | $10^{-6}$ | 9 | 2 |

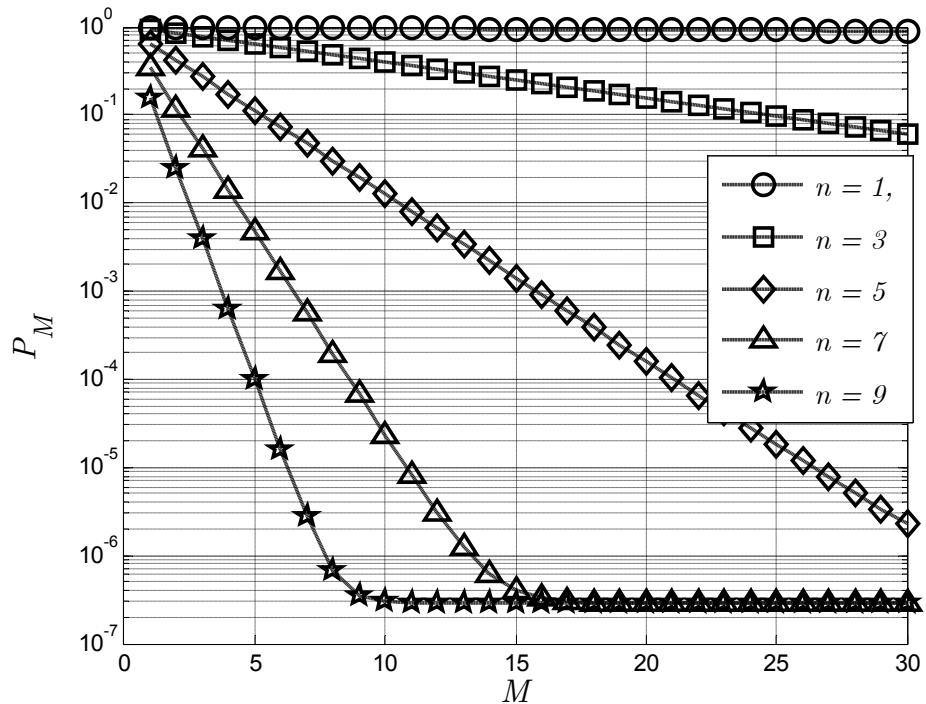

Fig. 4. $P_M$ vs. the concentration of SNMs (per 1000 $nm^3$), $M$, for all values of $n$.

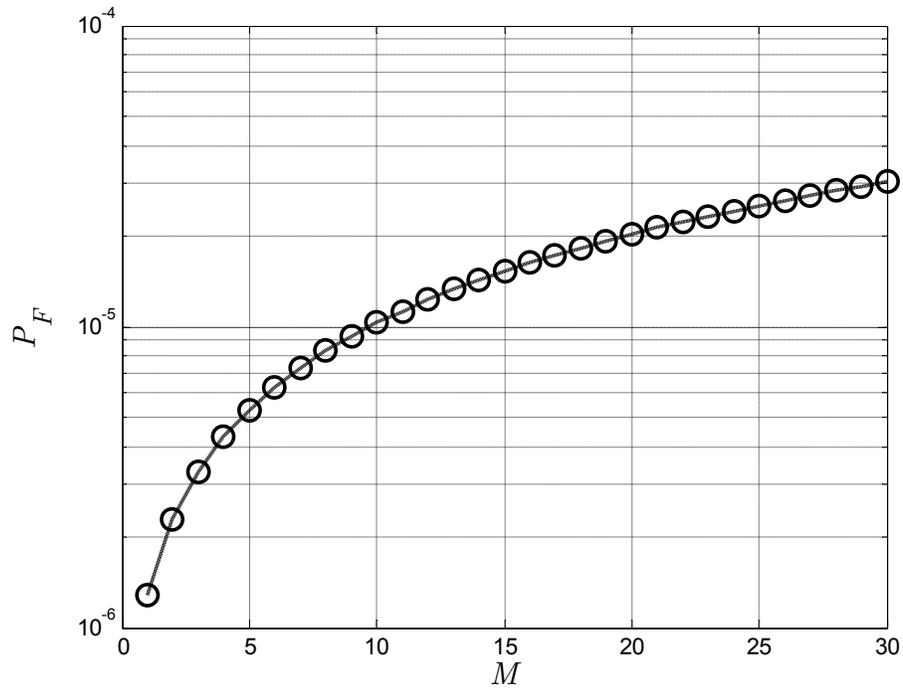

Fig. 5. $P_F$ vs. the concentration of SNMs (per 1000 $nm^3$), $M$, for all $n$.

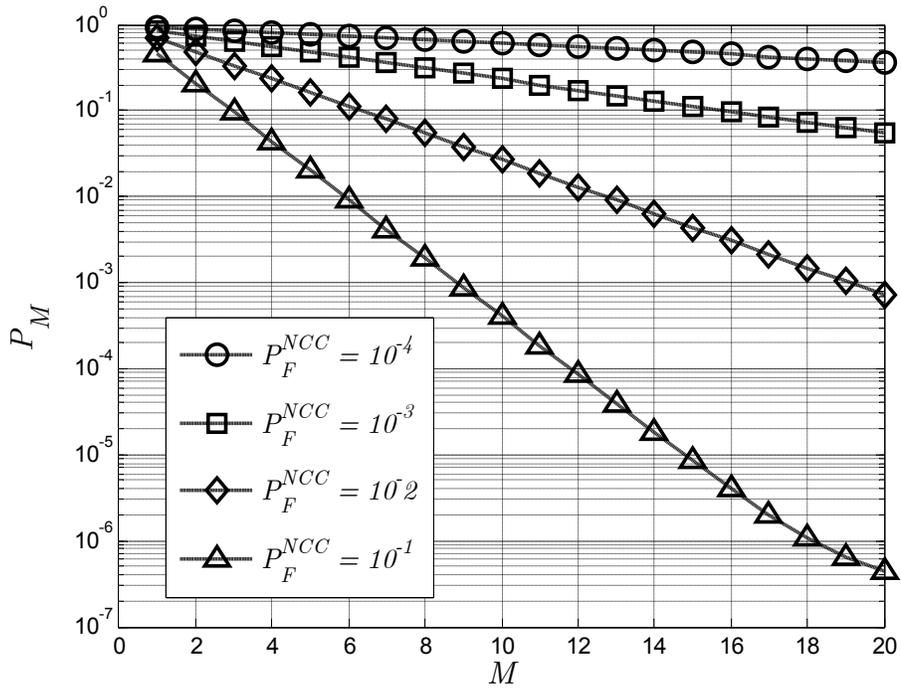

Fig. 6. $P_M$ vs. the concentration of SNMs, $M$ (per 1000 $nm^3$), for different values of $P_F^{NCC}$.

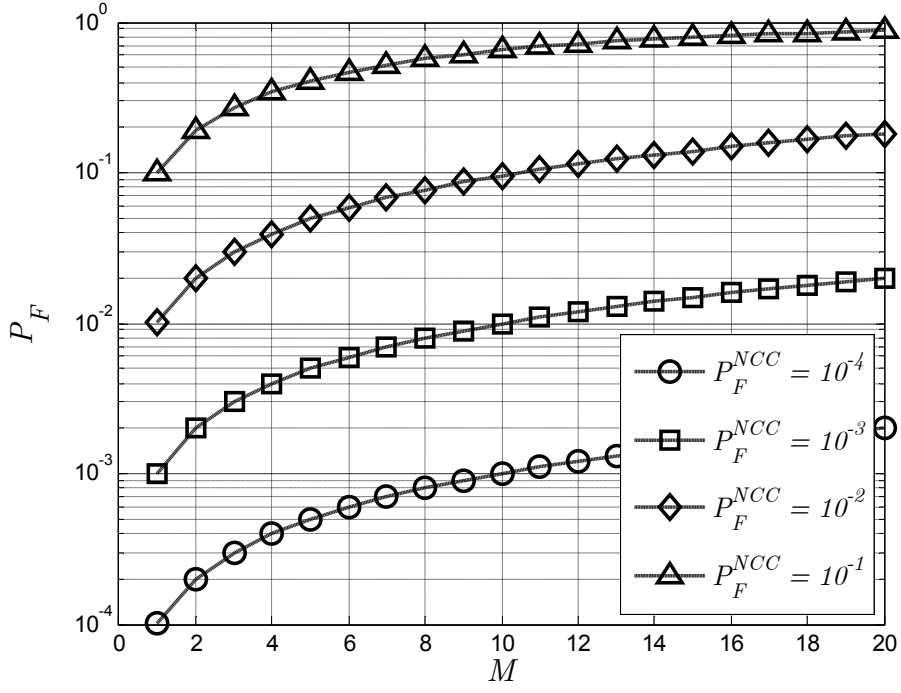

Fig. 7. $P_F$ vs. the concentration of SNMs, $M$ (per 1000 $nm^3$), for different values of $P_F^{NCC}$.

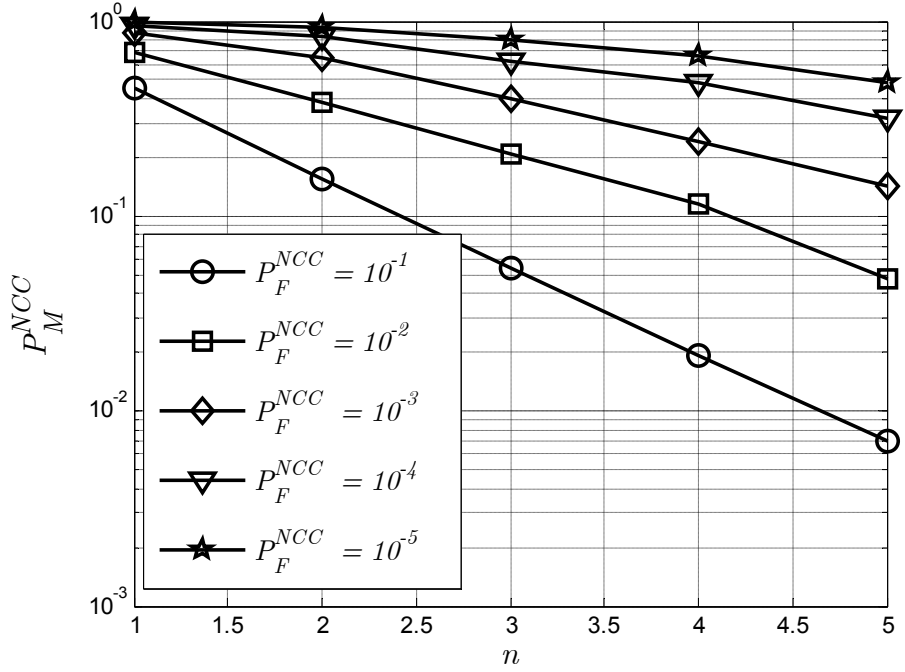

Fig. 8. $P_M^{NCC}$ vs. SNM observation time, $n$ for different values of $P_F^{NCC}$.

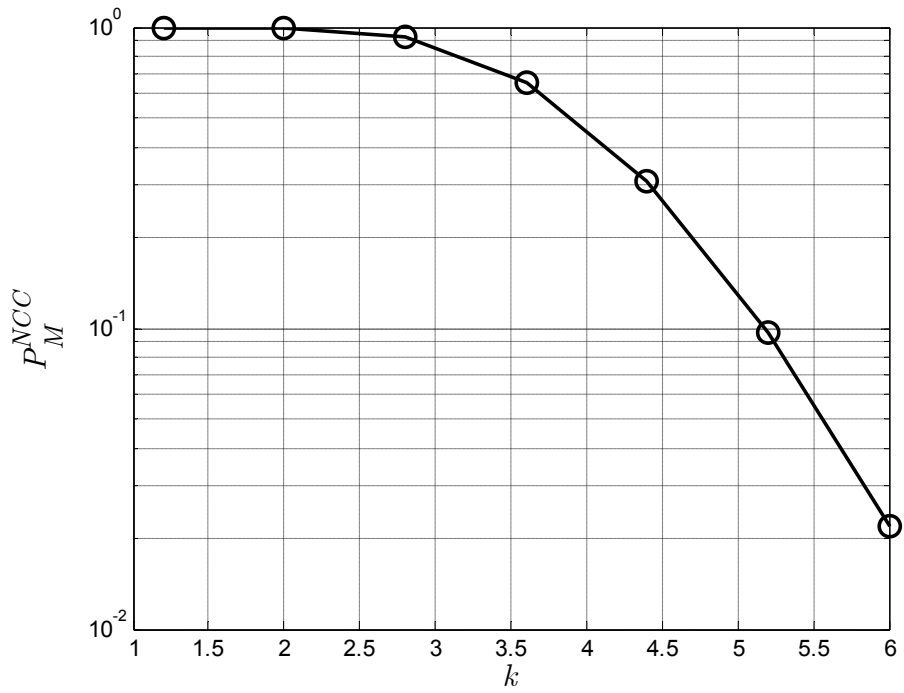

Fig. 9. $P_M^{NCC}$ vs. $k$

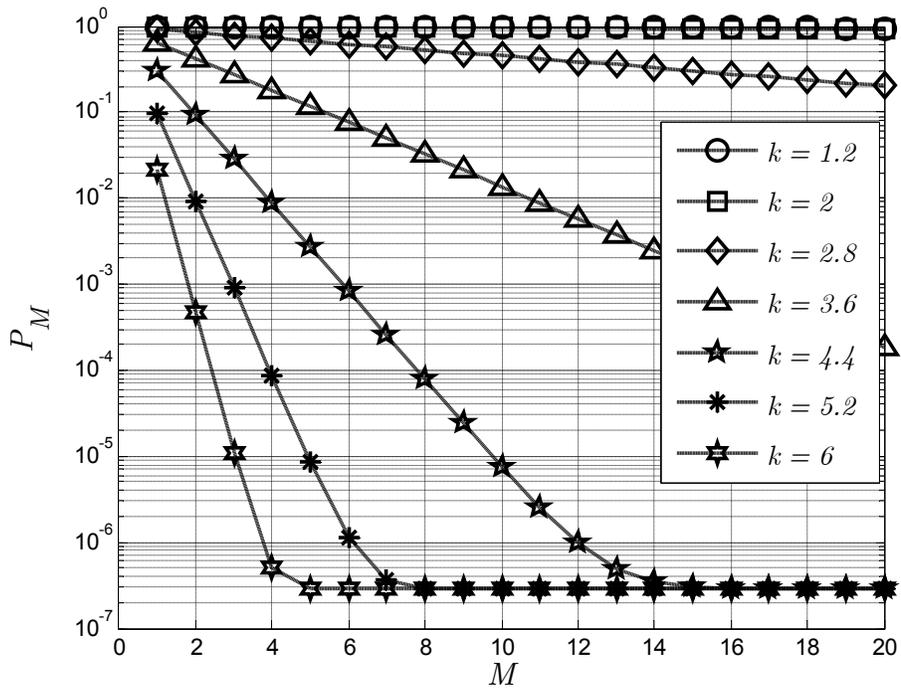

Fig. 10. $P_M$ vs. the concentration of SNMs, $M$ (per 1000 $nm^3$), for different values of $k$.

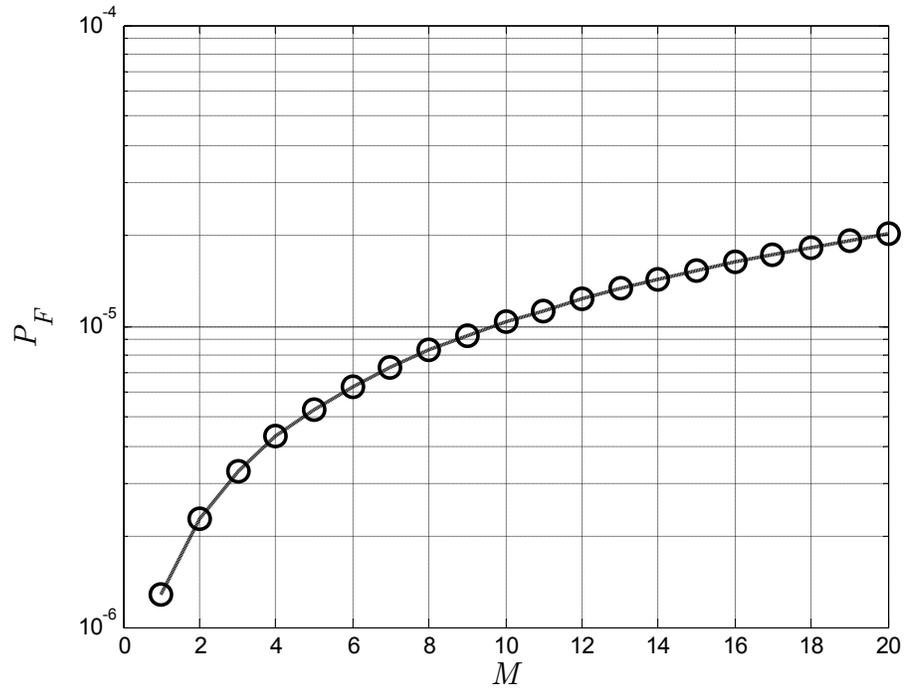

Fig. 11. $P_F$ vs. the concentration of SNMs, $M$ (per 1000 $nm^3$), for all values of $k$.

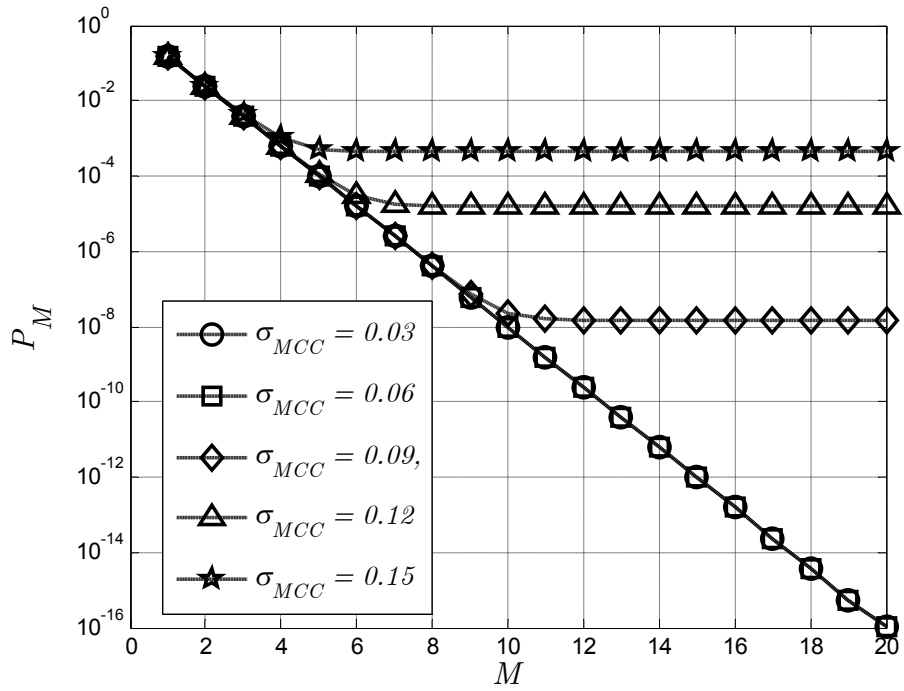

Fig. 12. $P_M$ vs. the concentration of SNMs, $M$ (per 1000 $nm^3$), for different values of $\sigma_{MCC}$.

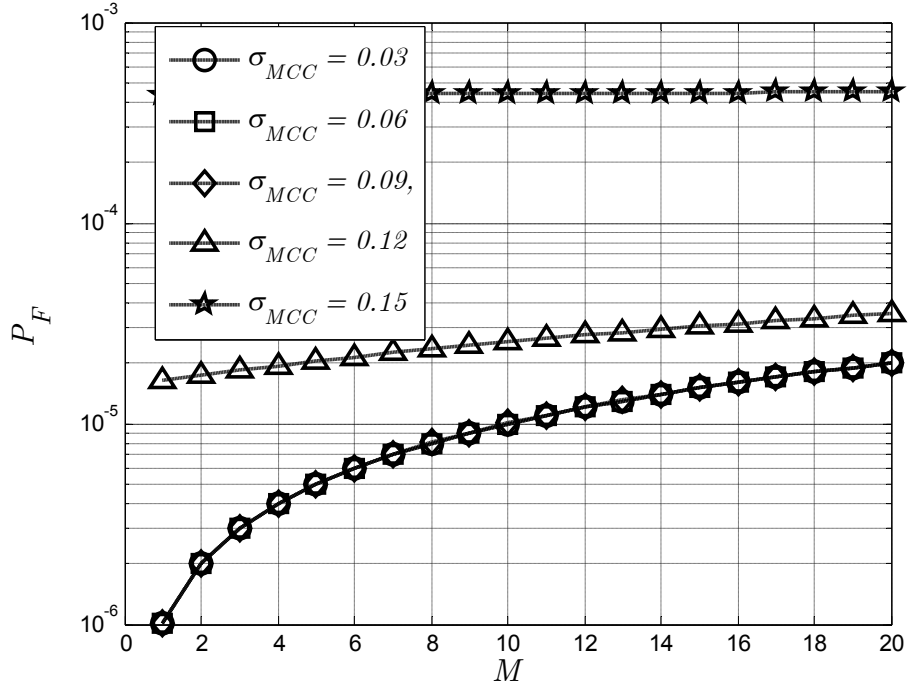

Fig. 13. $P_F$ vs. the concentration of SNMs, $M$ (per 1000 $nm^3$), for different values of $\sigma_{MCC}$.

## 6. Concluding remarks

A nano-scale abnormality detection scheme was proposed for the detection of competitor cells in a biomolecular nano-network. This is motivated by the early detection and classification of the diseases and their effective treatment. The proposed NADS is a two-tier network. The SNMs at the first tier act as receivers of an NCC modeling the molecular environment. The SNMs then communicate over a noisy channel to a DGN, which operates based on an OR fusion rule. The average number of received molecular bits $A$ per unit time serves as a feature for detecting the abnormalities at the SNMs. The detection performance of each SNM was analyzed with Poisson observations. Next, quantifying the overall NADS detection performance, optimized concentration of SNMs was obtained for a desired level of NADS reliability.

At the current stage of research into detection of diseases at the nano-scale, there are still many interesting related research problems. Here, we list a few of them:

- In this paper, the detection feature is set based on mathematical modeling and certain valid approximations. Developing more precise models or obtaining the exact detection feature based on experimental measurements in the target tissue is an interesting research avenue.

- Medical imaging is one approach to detection over the MCC. Realistic modeling of MCC noise is an interesting issue for enhancing the NADS performance for abnormality detection in the biomolecular environment.

- We opted a layered approach to the design of the considered two-tiered NADS. Such an approach is appealing as it lends itself to solutions with reasonable complexity. Alternatively, one may seek an optimum design of detectors for each of the NADS tiers based on the end-to-end NADS performance and the desired design problem in (16).

- Designing practical SNMs for detection of cancer or other diseases and taking the experimental constraints of those SNMs into consideration within the proposed NADS framework poses a number of other interesting and important research problems.

- The side effect of injected SNMs on the molecular environment plays an important role in the accuracy of the model and the performance of NADS. Hence, studying those effects is another key aspect of research in this field.

**Appendix A**

Based on the presented model for the NCC in Section 2, the number of existing molecules per unit time in (sample tissue of) volume size $V$ is $L_{ex,V} = L_{ex}V$. We define the event $X_j^i$, $X \in \{A,B\}$, $i \in \{C,P\}$, $j \in \{T,R\}$, where $\{C,P\}$ indicates current and previous time interval and $\{T,R\}$ indicates transmitting or receiving $X \in \{A,B\}$ over the NCC. Hence, $NP_H$ is given by

$$
\begin{aligned}
NP_H &= E\{NP|H_0\} \\
&= E\{NP|H_0, A_T^C\}\Pr\{A_T^C|H_0\} + E\{NP|H_0, B_T^C\}\Pr\{B_T^C|H_0\} \\
&= \left(E\{NP|H_0, A_T^C, A_R^P\}\Pr\{A_R^P|H_0, A_T^C\} + E\{NP|H_0, A_T^C, B_R^P\}\Pr\{B_R^P|H_0, A_T^C\}\right)\Pr\{A_T^C|H_0\} \\
&\quad + \left(E\{NP|H_0, B_T^C, A_R^P\}\Pr\{A_R^P|H_0, B_T^C\} + E\{NP|H_0, B_T^C, B_R^P\}\Pr\{B_R^P|H_0, B_T^C\}\right)\Pr\{B_T^C|H_0\} \\
&= E\{NP|H_0, A_T^C, A_R^P\}\Pr\{A_T^C, A_R^P|H_0\} + E\{NP|H_0, A_T^C, B_R^P\}\Pr\{A_T^C, B_R^P|H_0\} \\
&\quad + E\{NP|H_0, B_T^C, A_R^P\}\Pr\{B_T^C, A_R^P|H_0\} + E\{NP|H_0, B_T^C, B_R^P\}\Pr\{B_T^C, B_R^P|H_0\} \\
&= \left(E\{NP|H_0, A_T^C, A_R^P, A_R^C\}\Pr\{A_R^C|H_0, A_T^C, A_R^P\} \right.\\
&\quad \left. + E\{NP|H_0, A_T^C, A_R^P, B_R^C\}\Pr\{B_R^C|H_0, A_T^C, A_R^P\}\right)\Pr\{A_T^C, A_R^P|H_0\} \\
&\quad + \left(E\{NP|H_0, A_T^C, B_R^P, A_R^C\}\Pr\{A_R^C|H_0, A_T^C, B_R^P\} \right.\\
&\quad \left. + E\{NP|H_0, A_T^C, B_R^P, B_R^C\}\Pr\{B_R^C|H_0, A_T^C, B_R^P\}\right)\Pr\{A_T^C, B_R^P|H_0\} \\
&\quad + \left(E\{NP|H_0, B_T^C, A_R^P, A_R^C\}\Pr\{A_R^C|H_0, B_T^C, A_R^P\} \right.\\
&\quad \left. + E\{NP|H_0, B_T^C, A_R^P, B_R^C\}\Pr\{B_R^C|H_0, B_T^C, A_R^P\}\right)\Pr\{B_T^C, A_R^P|H_0\} \\
&\quad + \left(E\{NP|H_0, B_T^C, B_R^P, A_R^P\}\Pr\{A_R^P|H_0, B_T^C, B_R^P\} \right.\\
&\quad \left. + E\{NP|H_0, B_T^C, B_R^P, B_R^C\}\Pr\{B_R^C|H_0, B_T^C, B_R^P\}\right)\Pr\{B_T^C, B_R^P|H_0\}
\end{aligned}
\tag{36}
$$

where $H_0$ denotes a healthy molecular environment. The channel transition probabilities that appear in (36) can be expressed based on the notations in (13) and (14), e.g., $\Pr\{A_R^C|H_0, A_T^C, A_R^P\} = p_{AA}^{A_R^P}$. As the VTNM is considered a memoryless source, the events of the current transmission and the previously received molecular bit are independent. We also note that the channel is a stationary Markov channel in the steady state. Hence, the related terms in (36) can be simplified. For example,

$$\Pr\{A_T^C, A_R^P | H_0\} = \Pr\{A_T^C | H_0, A_R^P\} \Pr\{A_R^P | H_0\}$$
$$= \Pr\{A_T^C | H_0\} \Pr\{A_R^P | H_0\}$$
$$= P_A P_A', \tag{37}$$

where $P_A$ and $P_A'$ respectively denote the probabilities of transmitting and receiving molecular bit $A$ over the NCC. In the steady state regime, $P_A'$ is given by

$$P_A' = \Pr\{A_R^C | H_0\} = p_{AA}^{A_R^P} P_A \Pr\{A_R^C | H_0\} + p_{BA}^{A_R^P}(1-P_A)\Pr\{A_R^C | H_0\} + $$
$$p_{AA}^{B_R^P} P_A \left(1 - \Pr\{A_R^C | H_0\}\right) + p_{BA}^{B_R^P}(1-P_A)\left(1 - \Pr\{A_R^C | H_0\}\right) \tag{38}$$

By simplification, we have

$$P_A' = \frac{p_{AA}^{B_R^P} P_A + p_{BA}^{B_R^P}(1-P_A)}{1 - p_{AA}^{A_R^P} P_A - p_{BA}^{A_R^P}(1-P_A) + p_{AA}^{B_R^P} P_A + p_{BA}^{B_R^P}(1-P_A)}. \tag{39}$$

Using a similar approach, we have

$$P_B' = \frac{p_{AB}^{B_R^P} P_A + p_{BB}^{B_R^P}(1-P_A)}{1 - p_{AB}^{A_R^P} P_A - p_{BB}^{A_R^P}(1-P_A) + p_{AB}^{B_R^P} P_A + p_{BB}^{B_R^P}(1-P_A)} \tag{40}$$

Assuming a homogeneous molecular environment and considering a maximum of $N$ receptors for each SNM, the expectation terms in (36) can be obtained as

$$E\{NP | H_0, A_T^C, A_R^P, A_R^C\} = \min\left(\frac{L_{ex,V}}{S - N_A'}, N\right), \tag{41}$$

$$E\{NP | H_0, A_T^C, A_R^P, B_R^C\} = 0, \tag{42}$$

$$E\{NP | H_0, A_T^C, B_R^P, A_R^C\} = \min\left(\frac{L_{ex,V}}{S - N_B'}, N\right), \tag{43}$$

$$E\{NP|H_0, A_T^C, B_R^P, B_R^C\} = 0, \qquad (44)$$

$$E\{NP|H_0, B_T^C, A_R^P, A_R^C\} = \min\left(\frac{L_{ex,V}}{S - N_A'}, N\right), \qquad (45)$$

$$E\{NP|H_0, B_T^C, A_R^P, B_R^C\} = 0, \qquad (46)$$

$$E\{NP|H_0, B_T^C, B_R^P, A_R^C\} = \min\left(\frac{L_{ex,V}}{S - N_B'}, N\right), \qquad (47)$$

$$E\{NP|H_0, B_T^C, B_R^P, B_R^C\} = 0. \qquad (48)$$

Hence, incorporating equations (39), (40) and (41)-(48) into (36), $NP_H$ is derived as in (15).

**Appendix B**

The decision thresholds for NCC with Poisson observations are derived. The GLRT for hypothesis test of (17) is given by

$$GLRT: \frac{\max_{NP_R \neq NP_H} P(\mathbf{y}|NP_R)}{P(\mathbf{y}|NP_H)} = \frac{P(\mathbf{y}|\widehat{NP_R})}{P(\mathbf{y}|NP_H)} = \frac{\dfrac{e^{-n\widehat{NP_R}} \widehat{NP_R}^{\sum_{i=1}^{n} y_i}}{\prod_{i=1}^{n} y_i!}}{\dfrac{e^{-nNP_H} NP_H^{\sum_{i=1}^{n} y_i}}{\prod_{i=1}^{n} y_i!}} > \tau. \qquad (49)$$

where $\widehat{NP_R} = \arg\max_{NP_R}\left(P(\mathbf{y}|NP_R)\right)$, which simplifies to

$$e^{n(NP_H - \widehat{NP_R})} \frac{\widehat{NP_R}^{\sum_{i=1}^{n} y_i}}{NP_H^{\sum_{i=1}^{n} y_i}} > \tau. \qquad (50)$$

Taking the natural Logarithm from (50), we have

$$n(NP_H - \widehat{NP_R}) + \sum_{i=1}^{n} y_i \left(\log_e \widehat{NP_R} - \log_e NP_H\right) > \log_e \tau. \qquad (51)$$

Using $\sum_{i=1}^{n} y_i$ in (21), we obtain

$$n\left(NP_H - \widehat{NP_R}\right) + n\widehat{NP_R}\left(\log_e \widehat{NP_R} - \log_e NP_H\right) > \log_e \tau, \tag{52}$$

with some mathematical manipulation, we have

$$\widehat{NP_R}\left(\log_e \frac{\widehat{NP_R}}{NP_H} - 1\right) > \frac{\log_e \tau}{n} - NP_H. \tag{53}$$

Hence, the hypothesis test is simplified as

$$\begin{cases} H_0 & \widehat{NP_R}\left(\log_e \dfrac{\widehat{NP_R}}{NP_H} - 1\right) < \dfrac{\log_e \tau}{n} - NP_H \\[2ex] H_1 & \widehat{NP_R}\left(\log_e \dfrac{\widehat{NP_R}}{NP_H} - 1\right) > \dfrac{\log_e \tau}{n} - NP_H \end{cases} \tag{54}$$

Using (21) in (54), the desired result in (22) is derived. □

**Appendix C**

The SNM estimates the $NP_R$ by computing $1/n \sum_{i=1}^{n} y_i$ based on its observations and motivated by the LNN. If $NP_R$ deviates in a constrained range from $NP_H$, the molecular environment is considered healthy (hypothesis $H_0$). Otherwise, the SNM reports an abnormality (hypothesis $H_1$). On the other hand the mean of Poisson distribution is positive, hence a lower bound is set on $1/n \sum_{i=1}^{n} y_i$ in (24). To compute the threshold $\tau''$, we consider the false alarm probability as follows

$$\int_{\Lambda_1} P(\mathbf{y}|NP_R = NP_H) dy \leq \eta_1 \tag{55}$$

or equivalently

$$\int_{\Lambda_0} P(\mathbf{y}|NP_R = NP_H) dy \geq 1 - \eta_1, \tag{56}$$

in which $\Lambda_0$ and $\Lambda_1$, denote respectively the decision regions of $H_0$ and $H_1$. We have

$$\int_{NP_H-\tau''}^{NP_H+\tau''} P(\mathbf{y}|NP_R = NP_H)dy \geq 1-\eta_1, \tag{57}$$

which is expressed as

$$\Pr\left[NP_H - \tau'' \leq \frac{1}{n}\sum_{i=1}^{n} y_i \leq NP_H + \tau'' \middle| H_0\right] \geq 1-\eta_1. \tag{58}$$

If PDF of $y_i$, $\mathcal{P}(\lambda_i)$, is Poisson with mean $\lambda_i$, then the PDF of $\sum_{i=1}^{n} y_i$ is given by $\mathcal{P}\left(\sum_{i=1}^{n} \lambda_i\right)$ [30]. If $\lambda_i = \lambda$ $\forall i$, the PDF of $\sum_{i=1}^{n} y_i$ is then simply $\mathcal{P}(n\lambda)$. Therefore, (58) which is simplified as follows

$$\sum_{i=\max\{n(NP_H-\tau''),0\}}^{n(NP_H+\tau'')} e^{-nNP_H} \frac{(nNP_H)^i}{i!} \geq 1-\eta_1, \tag{59}$$

$$\sum_{i=0}^{n(NP_H+\tau'')} e^{-nNP_H} \frac{(nNP_H)^i}{i!} - \sum_{i=0}^{\max\{n(NP_H-\tau''),0\}-1} e^{-nNP_H} \frac{(nNP_H)^i}{i!} \geq 1-\eta_1, \tag{60}$$

$$\frac{\Gamma\left(\lfloor n(NP_H+\tau'')\rfloor+1, nNP_H\right)}{\lfloor n(NP_H+\tau'')\rfloor!} - \frac{\Gamma\left(\lfloor \max\{n(NP_H-\tau''),0\}-1\rfloor+1, nNP_H\right)}{\lfloor \max\{n(NP_H-\tau''),0\}-1\rfloor!} \geq 1-\eta_1. \tag{61}$$

In fact, (61) allows for numerical computation of $\tau''$, and hence, $P_D$ can be calculated as

$$P_D^{NCC} = \int_{\Lambda_1} P(\mathbf{y}|NP_R \neq NP_H)dy. \tag{62}$$

We have

$$P_D^{NCC} = \int_0^{\tau''-NP_H} P(\mathbf{y}|NP_R \neq NP_H)dy + \int_{\tau''+NP_H}^{\infty} P(\mathbf{y}|NP_R \neq NP_H)dy \tag{63}$$

$$\overset{(a)}{=} \Pr\left[\widehat{NP_R} < \tau'' - NP_H, \widehat{NP_R} > \tau'' + NP_H \middle| H_1\right] \tag{64}$$

$$\overset{(b)}{=} \Pr\left[\frac{1}{n}\sum_{i=1}^{n} y_i < \tau'' - NP_H, \frac{1}{n}\sum_{i=1}^{n} y_i > \tau'' + NP_H \middle| H_1\right] \tag{65}$$

$$\overset{(c)}{=} \sum_{i=0}^{\max(n(NP_H-\tau''),0)-1} e^{-n\widehat{NP_R}} \frac{(n\widehat{NP_R})^i}{i!} + \sum_{i=n(NP_H+\tau'')+1}^{\infty} e^{-n\widehat{NP_R}} \frac{(n\widehat{NP_R})^i}{i!} \tag{66}$$

$$=1+\frac{\Gamma\left(\lfloor\max(n(NP_H-\tau''),0)-1\rfloor+1,n\widehat{NP_R}\right)}{\lfloor\max(n(NP_H-\tau''),0)-1\rfloor!}-\frac{\Gamma\left(\lfloor n(NP_H+\tau'')\rfloor+1,n\widehat{NP_R}\right)}{\lfloor n(NP_H+\tau'')\rfloor!}. \quad (67)$$

in which (a) and (b) follow from (24) and (21), and (c) is written noting the Poisson distribution of $\sum_{i=1}^{n} y_i$. When a competitor cell is present in the NCC, $NP_R$ deviates from $NP_H$. Replacing (23) in (67), the desired result in (25) is derived. □

**Appendix D**

According to (28) and (29), $X_j$ may have two values of 0 or $G$. If $\sum_{j=1}^{M} X_j$ is less than $G$, it is equal to zero, hence (31) can be written as

$$\begin{cases} H_0 & \sum_{j=1}^{M} X_j = 0 \\ H_1 & \sum_{j=1}^{M} X_j \geq G. \end{cases} \quad (68)$$

At the DGN, we have a noisy observation of $\sum_{j=1}^{M} X_j$, which is denoted by $U$ as defined in (30). At the DGN, the purpose is to minimize the probability of error for the hypothesis test in (68). As the uniform cost [31] is to be calculated, the likelihood ratio is obtained as

$$\frac{p(U|H_1)}{p(U|H_0)} > 1 \quad (69)$$

$$\frac{\frac{1}{\sqrt{2\pi\sigma_{MCC}^2}} e^{-\frac{\left(U-\sum_{j=1}^{M} X_j\right)^2}{2\sigma_{MCC}^2}}}{\frac{1}{\sqrt{2\pi\sigma_{MCC}^2}} e^{-\frac{U^2}{2\sigma_{MCC}^2}}} > 1, \quad (70)$$

By simplification and taking the natural Logarithm, we have

$$-\frac{\left(U-\sum_{j=1}^{M}X_j\right)^2}{2\sigma_{MCC}^2}+\frac{U^2}{2\sigma_{MCC}^2}\overset{H_1}{>}0, \tag{71}$$

$$2U\left(\sum_{j=1}^{M}X_j\right)-\left(\sum_{j=1}^{M}X_j\right)^2\overset{H_1}{>}0, \tag{72}$$

$$U\overset{H_1}{>}0.5\sum_{j=1}^{M}X_j. \tag{73}$$

Hence, the probability of error for the MCC is given by

$$\begin{aligned}P_e &= \frac{1}{2}p(H_0|H_1)+\frac{1}{2}p(H_1|H_0),\\ &=\frac{1}{2}\left(1-P_D^{MCC}\right)+\frac{1}{2}P_F^{MCC},\end{aligned} \tag{74}$$

in which $P_D^{MCC}$ and $P_F^{MCC}$ are obtained as

$$\begin{aligned}P_D^{MCC} &= p(H_1|H_1),\\ &=\frac{1}{\sqrt{2\pi\sigma_{MCC}^2}}\int_{0.5\sum_{j=1}^{M}X_j}^{+\infty}e^{-\frac{\left(U-\sum_{j=1}^{M}X_j\right)^2}{2\sigma_{MCC}^2}}du,\\ &=1-Q\left(\frac{\sum_{j=1}^{M}X_j}{2\sigma_{MCC}}\right).\end{aligned} \tag{75}$$

$$\begin{aligned}P_F^{MCC} &= p(H_1|H_0),\\ &=\frac{1}{\sqrt{2\pi\sigma_{MCC}^2}}\int_{0.5\sum_{j=1}^{M}X_j}^{+\infty}e^{\frac{-U^2}{2\sigma_{MCC}^2}}du,\\ &=Q\left(\frac{\sum_{j=1}^{M}X_j}{2\sigma_{MCC}}\right).\end{aligned} \tag{76}$$

For hypothesis $H_1$, the minimum of $\sum_{j=1}^{M}X_j$ is $G$, which is associated to the maximum probability of error.

Considering $\sum_{j=1}^{M}X_j=G$ in (73), (75) and (76) the desired results are obtained. Of course, this leads to a larger error probability in comparison to the case with the optimal detector. Our experiments reveal that this upper bound is tight when the signal to noise ratio in MCC is higher than a specified level of about 10 dB. □